\documentclass[12pt]{article}

\usepackage{latexsym}
\usepackage{graphics}
\newcommand{\be}{\begin{equation}}
\newcommand{\ee}{\end{equation}}
\newcommand{\ben}{\begin{eqnarray}}
\newcommand{\een}{\end{eqnarray}}
\newcommand{\bea}{\begin{eqnarray}}
\newcommand{\eea}{\end{eqnarray}}

\begin{document}

\setlength{\baselineskip}{19pt}
\title{
\normalsize
\mbox{ }\hspace{\fill}
\begin{minipage}{7cm}
{\tt }{\hfill}
\end{minipage}\\[5ex]
{\large\bf Anomaly-free representations of the holonomy-flux algebra
 \\[1ex]}}
\author{SangChul Yoon
\\
\\
{\it 128-6 Jongro-Ku Gye-Dong},\\
{\it Seoul 110-270, Korea}\\
}

\maketitle

\thispagestyle{empty}

\begin{abstract}
We work on the uniqueness \cite{LOST} of representations of the
holonomy-flux algebra in loop quantum gravity. We argue that for
analytic diffeomorphisms, the flux operators can be only constants
as functions on the configuration space in representations with no
anomaly, which are zero in the standard representation.
\end{abstract}

In loop quantum gravity\footnote{Among many nice introductions,
\cite{Ma} is good for understanding the uniqueness of the
holonomy-flux algebra.}, the configuration variables are holonomies
$h_e[A]$ of a connection field
 and the momentum variables are surface integrals $E(S,f)$ of a
triad field. Quite interestingly, the Poisson brackets between the
momentum variables do not vanish. The origin of  this
non-commutativity comes from the two-dimensional singular smearing
of $E(S,f)$ \cite{ACZ} and $E(S,f)$ can be understood as some vector
fields $X(S,f)$ on the configuration space $\mathcal{A}$. In the
standard representation, every holonomy operator is multiplication
and every flux operator is derivation on the Hilbert space $L^2(
\bar {\mathcal{A}} , \mu) $ \cite{ALMMT}.

Representations  of the holonomy-flux algebra  were further
investigated in \cite{Sahlmann}. It was motivated by the fact that
the momentum variables $E(S,f)$ are not constants\footnote{The
Poisson brackets between two momentum functions are zero.} on the
configuration space $\mathcal{A}$ and proposed that $E(S,f)$ can be
functions
 $F(S,f)$ on $\mathcal{A}$

\bea  \pi(E(S,f)) = X(S,f) + F(S,f) \eea where a map  $\pi$ is a
representation of the holonomy-flux algebra. Later it was found that
$F(S,f)$ are real valued assuming that $\pi$ is covariant with
respect to the group of the analytic diffeomorphisms \cite{LO1}.
Finally it was shown that $F(S,f)$ vanish by proving that their
norms which are obtained from a state via the GNS construction are
zero requesting that they are invariant under the group of the
semianalytic diffeomorphisms \cite{LOST}.

Whether this uniqueness holds for the analytic diffeomorphisms is an
open question. In this paper, we argue that we only have freedom to
take some constants besides zero as $F(S,f)$ for the analytic
diffeomorphisms provided that there is no anomaly in the
representations of the holonomy-flux algebra. Considering the
Poisson bracket between $E(S_1,f_1)$ and $E(S_2,f_2)$ is zero if
$S_1$ and $S_2$ are disjoint, we request that

\bea \pi ([E(S_1,f_1),E(S_2,f_2)])=0 \qquad\mbox{for}\qquad S_1 \cap
S_2=\phi. \eea  In a GNS representation, it becomes \bea
X(S_1,f_1)(F(S_2,f_2))-X(S_2,f_2)(F(S_1,f_1))=0
\qquad\mbox{for}\qquad S_1 \cap S_2=\phi. \eea From the definition
of $E(S,f)$, we request that \bea F(S_1 \cup
S_2,f)=F(S_1,f)+F(S_2,f)-F(S_1 \cap S_2,f). \eea We do not consider
some difficulty with boundaries which is not essential in our
purpose or we only consider surfaces which do not include their
boundaries. \\\\
{\bf Theorem 1.}  {\it Suppose $\pi$ is a representation of the
holonomy-flux algebra and is covariant with respect to the group of
the analytic diffeomorphisms. Suppose also, there is no anomaly in
the representation. Then, all the $L^2(\bar{\mathcal{A}},\mu)$
functions $F(S,f)$ are constants on the configuration space
$\mathcal{A}$.}
\\

To show this, we need two properties of Hilbert
spaces \cite{Conway}. \\ \\
{\bf Theorem 2.} {\it For any vector $h$ and a given basis \{$e_i$\}
in a Hilbert space $H$, the expansion $h=\sum_{i} a_i e_i$ is
unique.}\\\\
{\bf Theorem 3.} {\it For any vector $h$ and a given basis \{$e_i$\}
in a Hilbert space $H$, $< h , e >$ $ \neq 0$ for at most a
countable number of vectors $e$ in \{$e_i$\}. } \\ \\ Theorem 2 is
obvious and Theorem 3 comes from the condition that any vector has
finite norm.

We are going to show that when we express $F(S,f)$ in terms of the
spin network basis \cite{RS}\cite{Baez}, all the edges are trivial.
\bea F(S,f)=\sum_{(\alpha,j)} C_{\alpha,j} \phi_{\alpha,j} \eea
where $\alpha$ is a graph and $j$ assigns to each edge of $\alpha$ a
non-trivial irreducible representations of SU(2). Because holonomy's
dependence on a connection disappears for a trivial edge, Theorem 1 will
be proved in this case.\\\\
{\bf Lemma 4.} {\it Any edge $e$ in $\alpha$ lies on $S$.} \\\\
{\it Proof.}  Assume a analytic diffeomorphism $\psi$ with $ \psi(S)
\cap S = 0$. Take $x \in e$ and $\psi(x) = S  \cap \psi(e)$, where
$e$ is an edge in a graph $\alpha$ of (5) and $e \not\subset S$. Now
consider $E(S,f)$ and $E(\psi(S),(\psi^{-1})^*(f))$. Because $
\psi(S) \cap S = 0$, $ \pi ([E(S,f),
E(\psi(S),(\psi^{-1})^*(f))])=0$. Therefore there should be a graph
in $\{\alpha\}$ which includes the edge $\psi(e)$ to satisfy (3).
Because $\psi$ is arbitrary, $F(S,f)$ should be expanded with an
uncountable number of graphs. By Theorem 3, $F(S,f)$ can not be
$L^2(\bar{\mathcal{A}},\mu)$. Therefore $e \subset S$.\\\\
{\bf Lemma 5.} {\it To satisfy (4), any edge $e$ in $\alpha$ is
trivial.}\\\\
{\it Proof.} By Lemma 4, $e$ lies on $S$. Suppose $S_1$ and $S_2$ in
$S$. Assume $e \subset S_1 \cup S_2$ with $e \not\subset S_1$ and $e
\not\subset S_2$. We can see that $F(S_1,f)$, $F(S_2,f)$ and $F(S_1
\cap S_2,f)$ do not have a component containing $e$ by Lemma 4..
Therefore (4) can not be satisfied by Theorem 2. It is possible only when $e$ is trivial.\\\\

We thank J. Lee and J. Yoon for useful discussions.\\\\ \\\\\\\\\\



\begin{thebibliography}{100}
\bibitem{LOST}J. Lewandowski, A. Okolow, H. Sahlmann and T. Thiemann, Uniqueness of diffeomorphism
invariant states on holonomy-flux algebras, {\it Commun. Math. Phys.
} {\bf 267}, 703-733 (2006), gr-qc/0504147
\bibitem{Ma}M. Han, W. Huang and Y. Ma, Fundamental structure of loop quantum gravity, {\it Int. J. Mod.
Phys.} {\bf D16}, 1397-1474 (2007), gr-qc/0509064
\bibitem{ACZ}A. Ashtekar, A. Corichi and J. A. Zapata, Quantum theory of geometry III:
Noncommutativity of Riemannian structures, {\it Class. Quant. Grav.}
{\bf 15}, 2955-2972 (1998), gr-qc/9806041
\bibitem{ALMMT}A. Ashtekar, J. Lewandowski, D. Marolf, J. Mourao and
T. Thiemann, Quantization of diffeomorphism invariant theories of
connections with local degrees of freedom, {\it J. Math. Phys.} {\bf
36}, 6456-6493 (1995), gr-qc/9504018
\bibitem{Sahlmann}H. Sahlmann, Some comments on the representation theory of the
algebra underlying loop quantum gravity, gr-qc/0207111
\bibitem{LO1}J. Lewandowski and A. Okolow, Diffeomorphism covariant representations
of the holonomy flux $\ast$ algebra, {\it Class. Quant. Grav.} {\bf
20}, 3543-3568 (2003), gr-qc/0302059

\bibitem{Conway}J. B. Conway, {\it A Course in Functional Analysis},
Springer-Verlag, New York 1990
\bibitem{RS}C. Rovelli and L. Smolin, Spin networks and quantum
gravity, {\it Phys. Rev.} {\bf D52}, 5743-5759 (1995), gr-qc/9505006
\bibitem{Baez}J. C. Baez, Spin network states in gauge theory,
{\it  Adv. Math. } {\bf 117}, 253-272 (1996), gr-qc/9411007
\end{thebibliography}
\end{document}